\documentclass[prl,aps,reprint,noshowpacs]{revtex4-1}
\usepackage[utf8x]{inputenc}
\usepackage{amsmath,amssymb,graphicx}
\newcommand{\avg}[1]{\langle #1\rangle}
\newcommand{\dd}{\mathrm{d}}
\newcommand{\ee}{\mathrm{e}}
\newcommand{\onefig}[1]{\includegraphics[scale=0.28]{#1}}
\begin{document}
\title{Temporal effects in the growth of networks}
\author{Matúš Medo, Giulio Cimini, Stanislao Gualdi}
\affiliation{Physics Department, University of Fribourg,
CH-1700 Fribourg, Switzerland}
\date{\today}
\begin{abstract}
We show that to explain the growth of the citation network by
preferential attachment (PA), one has to accept that individual
nodes exhibit heterogeneous fitness values that decay with time.
While previous PA-based models assumed either heterogeneity or
decay in isolation, we propose a simple analytically treatable
model that combines these two factors. Depending on the input
assumptions, the resulting degree distribution shows an
exponential, log-normal or power-law decay, which makes the
model an apt candidate for modeling a wide range of real
systems.
\end{abstract}

\maketitle

Over the years, models with preferential attachment (PA) were
independently proposed to explain the distribution of the number
of species in a genus \cite{Yule25}, the power-law distribution
of the number of citations received by scientific papers
\cite{Price76} and the number of links pointing to WWW pages
\cite{BarAlb99}. A theoretical description of this class of
processes and the observation that they generally lead to
power-law distributions are due to Simon \cite{Simon55}.
Notably, the application of PA to WWW data by Barabási and
Albert helped to initiate the lively field of complex networks
\cite{Newman03}. Their network model, which stands at the center
of attention of this work, was much studied and generalized to
include effects such as presence of purely random connections
\cite{Liu02}, non-linear dependence on the degree
\cite{KrReLe00}, node fitness \cite{BiaBar01} and others
\cite[Ch. 8]{AlbBar02}.

Despite its success in providing a common roof for many
theoretical models and empirical data sets, preferential
attachment is still little developed to take into account the
temporal effects of network growth. For example, it predicts a
strong relation between a node's age and its degree. 
While such first-mover advantage~\cite{Newman09} plays a
fundamental role for the emergence of scale free topologies in
the model, it is  a rather unrealistic feature for several real
systems (\emph{e.g.}, it is entirely absent in the
WWW~\cite{AdaHub00} and significant deviations are found in
citation data~\cite{Newman09,Redner05}). This motivates us to
study a model of a growing network where a broad degree
distribution does not result from strong time bias in the system.
To this end we assign fitness to each node and assume that this
fitness decays with time---we refer it as relevance henceforth.
Instead of simply classifying the vertices as active or
inactive, as done in~\cite{AmScBaSt00,LeJaLa05}, we use real
data to investigate the relevance distribution and decay therein
and build a model where decaying and heterogeneous relevance are
combined.

Models with decaying fitness values (``aging'') were shown to
produce narrow degree distributions (except for very slow
decay)~\cite{DoMe00} and widely distributed fitness values were
shown to produce extremely broad distributions or even a
condensation phenomenon where a single node attracts a
macroscopic fraction of all links~\cite{BiaBar01b}. We show that
when these two effects act together, they produce various
classes of behavior, many of which are compatible with
structures observed in real data sets.

Before specifying a model and attempting to solve it, we turn to
data to provide support for our hypothesis of decaying
relevance. We use here the citation data provided by the
American Physical Society (APS) which contains all 450\,084
papers published by the APS from 1893 to 2009 together with
their 4\,691\,938 citations of other papers from APS journals.
It is particularly fitting to use citation data for our work
because ordinary PA with direct proportionality to the node
degree was detected in this case by previous
works~\cite{Newman09,JeNeBa03}. Data analysis according
to~\cite{CSN09} reveals that the best power-law fit to the
in-degree data has lower bound $k_{\mathrm{min}}=50$ and
exponent $2.79\pm0.01$. Though $p$-values greater than $0.10$
are only achieved for $k_{\mathrm{min}}\gtrsim150$, log-normal
distribution does not appear to fit the data particularly
better. Since PA can be best imagined to model citations within
one field of research, we consider in our analysis also a subset
of papers about the theory of networks. We identify them using
the PACS number 89.75.Hc (``Networks and genealogical
trees'')---in this way we obtain a small data set with 985
papers and 4\,395 citations among them.

Denoting the in-degree of paper $i$ at time $t$ as $k_i(t)$ and
assuming that during next $\Delta t$ days, $C(t,\Delta t)$ new
citations are added to papers in the network, preferential
attachment predicts that the number of citations received by
paper $i$ is
$\Delta k_i(t,\Delta t)_{PA}=C(t,\Delta t)k_i(t)/\sum_j k_j(t)$.
If in reality, $\Delta k_i(t,\Delta t)$ citations are received,
the ratio between this number and the expected number of
received citations defines the paper's relevance
\begin{equation}
\label{X-def}
X_i(t,\Delta t):=\frac{\Delta k_i(t,\Delta t)\sum_j k_j(t)}
{C(t,\Delta t)k_i(t)}.
\end{equation}
This expression is obviously undefined for $k_i(t)=0$ which
stems from the known limitation of the PA in requiring an
additional attractiveness factor to allow new papers to gain
their first citation. Although one could try to include this
effect in our analysis, we simply compute $X_i(t,\Delta t)$ only
when $k_i(t)\geq1$. Similarly we exclude time periods when no
citations are given and $C(t,\Delta t)=0$.

\begin{figure}
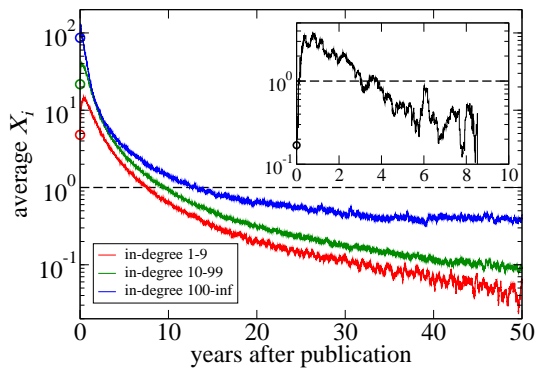

\centering
\onefig{X_ratio-91}
\caption{Time decay of the average relevance values (based on
$\Delta t=91\,\mathrm{days}$) for papers divided into groups
according to their final in-degree (color online). The dashed
line shows $X=1$ indicating exact preferential attachment and
open circles show the initial relevance values. The inset shows
results for papers about the theory of networks.}
\label{fig:X_ratio}
\end{figure}

\begin{figure}
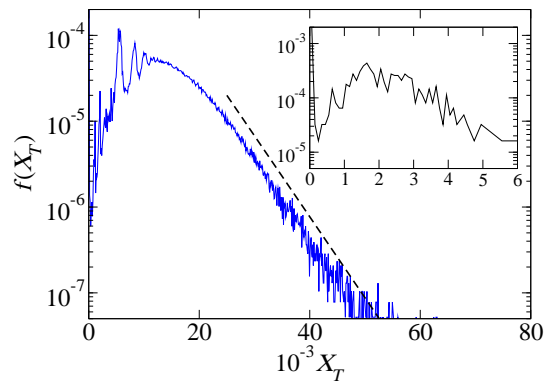

\centering
\onefig{X_tot-91}
\caption{The distribution of the total relevance $X_T$ in the
studied data (color online). For $X_T\gtrsim25\cdot 10^3$,
$f(X_T)$ decays as $\exp[-\alpha X_T]$ with
$\alpha=(21.7\pm0.2)\cdot 10^{-3}$ (denoted with the indicative
dashed line). The peak at $X_T=0$ is due to approximately
$60\,000$ papers without citations. The inset shows results for
papers about the theory of networks.}
\label{fig:X_tot}
\end{figure}

Figure~\ref{fig:X_ratio} shows how the average relevance of
papers with different final in-degree values decays with time
after their publication. We see that the relevance values indeed
decay and this decay is initially very fast (for papers with the
highest final in-degree, it is by a factor of 100 in less than
three years). However, the exponential decay reported in
\cite{Zhu03} appears to have only very limited validity (up to
five years after the publication date). After 10 or more years,
the decay becomes very slow or even vanishes, producing a
stationary relevance value $r_0$. Figure~\ref{fig:X_tot} depicts
the distribution of the total relevance $X_T(i)=\sum_t X_i(t)$
and shows that, perhaps contrary to one's expectations, this
distribution is rather narrow with an exponential decay for
$X_T\gtrsim 25\cdot 10^3$. An exponential-like tail appears also
when the analysis is restricted to papers of a similar age which
means that it is not only an artifact of the papers' age
distribution. One could attempt to fit this data with, for
example, a Weibull distribution as in~\cite{BoMaGo04}. We shall
see later that it is the tail behavior of $X_T$ what determines
the tail behavior of the degree distribution, hence the current
level of detail suffices our purpose. We can conclude that in
the studied citation data, relevance values exhibit time decay
and papers' total relevances are rather homogeneously
distributed, showing an exponential decay in the tail.

Now we proceed to a model based on the above-reported empirical
observations. We consider a uniformly growing undirected
network which initially consists of two connected nodes. At time
$t$, a new node is introduced and linked to an existing node $i$
where the probability of choosing node $i$ is
\begin{equation}
\label{basic}
P(i,t)=\frac{k_i(t) R_i(t)}{\sum_{j=1}^t k_j(t) R_j(t)}
\end{equation}
which has the same structure as assumed before in
\cite{DoMe00,Zhu03}. Here $k_j(t)$ and $R_j(t)$ is degree and
relevance of node $j$ at time $t$, respectively~\cite{footnote1}.
Our goal is to determine whether a stationary degree distribution
exists and find its functional form.

Eq.~(\ref{basic}) represents a complicated system where
evolution of each node's degree depends not only on the node
itself but also on the current degrees and relevances of all
other nodes. The key simplification is based on the assumption
that at any time moment (except for a short initial period),
there are many nodes with non-negligible values of
$k_i(t)R_i(t)$. The denominator of Eq.~(\ref{basic}) is then a
sum over many contributing terms and therefore it fluctuates
little with time. This allows us to approximate the exact
selection probability $P(i,t)$ with
\begin{equation}
\label{simpl}
P(i,t)=\frac{k_i(t) R_i(t)}{\Omega(t)}
\end{equation}
where $\Omega(t)$ is now just a normalization factor.

\begin{figure}
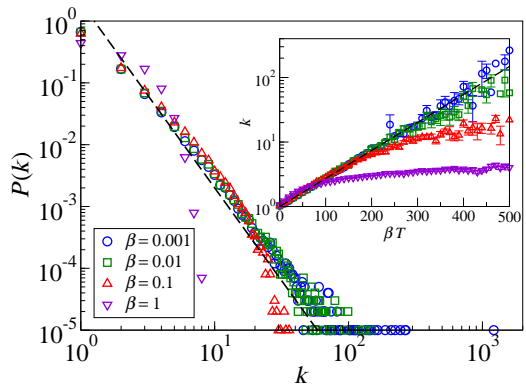

\centering
\onefig{Pk_numerical}
\caption{Simulation results for the studied model where
$R_i(t)=R_i(0)\,\ee^{-\beta(t-t_i)}$ ($t_i$ is the time when node
$i$ entered the network), $R_i(0)$ values are drawn from an
exponential distribution and the final number of nodes is $10^5$
(color online). Since the decay is the same for all nodes,
distributions of $R_i(0)$ and $T_i$ have the same functional
form. The indicative dashed line has the slope of $-3$. The
inset shows the dependency between $\beta T$ and the average
node degree; the dashed line follows from Eq.~(\ref{k^F}).}
\label{fig:simulations}
\end{figure}

If $R_i(t)$ decays sufficiently fast (faster than $1/t$) and
$\lim_{t\to\infty} R_i(t)=0$, the initial growth of $\Omega(t)$
stabilizes at a certain value $\Omega^*$ which shall be
determined later by the requirement of self-consistency.
The master equation for the degree distribution $p(k_i,t)$
now has the form $p(k_i,t+1)=(1-k_iR_i(t)/\Omega^*)p(k_i,t)+
(k_i-1)R_i(t)p(k_i-1,t)/\Omega^*$. Note that the stationarity of
$p(k_i,t)$ in our case is due to transition probabilities that
vanish because $\lim_{t\to\infty}R_i(t)=0$. Before tackling the
degree distribution itself, we examine the expected final degree
of node $i$, $\avg{k_i^F}$. By multiplying the master equation
with $k_i$ and summing it over all $k_i$, we obtain a difference
equation $\avg{k_i(t+1)}=\avg{k_i(t)}\big(1+R_i(t)/\Omega^*\big)$.
If $R_i(t)$ decays sufficiently slowly, we can switch to
continuous time to obtain
$\dd\avg{k_i(t)}/\dd t=k_i(t)R_i(t)/\Omega^*$ which together
with $\avg{k_i(t_i)}=1$ yields
\begin{equation}
\label{k^F}
\avg{k_i^F}=\exp\Big(\frac1{\Omega^*}
\int_{t_i}^{\infty}\! R_i(t)\,\dd t\Big).
\end{equation}
Here $t_i$ is the time when node $i$ is introduced to the
system (in our case, $t_i=i$).
When the continuum approximation is valid, this result is well
confirmed by numerical simulations (see the inset in
Fig.~\ref{fig:simulations}). To observe saturation of the degree
growth for an infinitely growing network, the total relevance
$T_i:=\int_{t_i}^{\infty} R_i(t)\,\dd t$ must be finite and hence
$R_i(t)$ must decay faster than $1/t$ for all nodes. To assess
the error of the continuum approximation, one can use the
Taylor expansion to write
$\avg{k_i(t+1)}-\avg{k_i(t)}\approx\dd\avg{k_i(t)}/\dd t+
\tfrac12\dd^2\avg{k_i(t)}/\dd t^2$. The second derivative term
can be approximately evaluated using Eq.~(\ref{k^F}) and it can
be shown that it's negligible when $\lvert\dot R_i(t)\rvert\ll
R_i(t)$, which is consistent with our initial assumption
that $R_i(t)$ decays sufficiently slowly for all $i$.

Since $\Omega^*$ is the same for all nodes, Eq.~(\ref{k^F})
demonstrates that a node's expected final degree depends only on
its total relevance $T_i$. Therefore we can use the continuum
approach to compute $\Omega^*$ directly from its definition as
$\Omega^*=\int\varrho(T)\avg{\Omega(T)}\,\dd T$ where
$\avg{\Omega(T)}\approx\lim_{t\rightarrow\infty}\int_0^tR(t-t_0)
\avg{k(t-t_0)}\,\dd t_0=\int_0^{\infty}R(t)\avg{k(t)}\,\dd t=
\Omega^*\big(\ee^{T/\Omega^*}-1\big)$, as there is only one node
with total relevance $T$ which contributes to $\avg{\Omega(T)}$
with $R(t)\avg{k(t)}$ for each $t$. When $\varrho(T)$ is given,
the resulting equation
\begin{equation}
\label{omega*}
\int\varrho(T)\,\ee^{T/\Omega^*}\dd T=2
\end{equation}
can be used to find $\Omega^*$. Alternatively, the construction
constraint of the average network's degree in the large time
limit, $\avg{k}=2$, implies
$\int_0^{\infty} \varrho(T)\avg{k^F(T)}\,\dd T=2$ which gives
the same equation for $\Omega^*$. Note that when $\varrho(T)$
decays slower than exponentially, the integral in
Eq.~(\ref{omega*}) diverges and no $\Omega^*$ can satisfy the
system's requirements, implying that in this case no stationary
value of $\Omega^*$ is established.

Similarly to $\avg{k_i(t)}$, degree fluctuations for
nodes of a given total relevance can be derived from the master
equation. When $\lvert\dot R_i(t)\rvert\ll R_i(t)$, the
continuum approximation can be again shown to be valid and
yields
\begin{equation}
\label{dk2}
\dd\avg{k_i^2}/\dd t=
R_i(t)\big(\avg{k_i(t)}+2\avg{k_i^2(t)}\big)/\Omega^*
\end{equation}
where $\avg{k_i^2(0)}=1$ and which can be solved for general
$R_i(t)$ to obtain the stationary standard deviation of the
node's degree
\begin{equation}
\label{sigma_k}
\sigma_k(T_i)=\big(\ee^{2T_i/\Omega^*}-\ee^{T_i/\Omega^*}
\big)^{1/2}.
\end{equation}
When $T_i=T$ for all nodes, Eq.~(\ref{omega*}) implies
$\ee^{T/\Omega^*}=2$ and therefore $\sigma_k=\sqrt{2}$. We see
that the resulting degree distribution $f(k)$ is very narrow
which is not the case in most real complex networks. One has to
proceed to heterogeneous $T_i$ values.

Since the distribution $f(k_i\vert T_i)$ is very narrow, one can
use the distribution $\varrho(T)$ together with Eq.~(\ref{k^F})
and $f(k)\,\dd k=\varrho(T)\,\dd T$ to obtain the degree
distribution $f(k)$. If $T_i$ are drawn from a distribution with
finite support, the support of $f(k)$ is also finite which is not
of interest for us (though it may be appropriate to model
some systems). If $T_i$ follow a truncated normal distribution
(the truncation is needed to ensure $T_i\geq0$ and
$\avg{k_i}\geq1$), it follows immediately that $f(k)$ is
log-normally distributed which may be of great relevance in many
cases~\cite{CSN09,Mitz04}. We finally consider $T_i$ values that
follow a fast-decaying exponential distribution
$\varrho(T)=\alpha\exp[-\alpha T]$ which is supported by the
analysis of citation data presented in Figure~\ref{fig:X_tot}.
By transforming from $\varrho(T)$ to $f(k)$, we then obtain
$f(k)\sim k^{-1-\alpha\Omega^*}$. From Eq.~(\ref{omega*}) it
follows that in this case is $\Omega^*=2/\alpha$, hence the
power-law exponent is $\gamma=3$. We see that even a very
constrained exponential distribution of $T$ leads to a
scale-free distribution of node degree---the exponent of this
distribution is in fact the same as in the original PA model. As
shown in Fig.~\ref{fig:simulations}, numerical simulations
confirm that this result truly realizes in a wide range of
parameters.

Motivated by Fig.~\ref{fig:X_tot}, one may ask what happens when
$T$ is exponentially distributed only in its tail. We take a
simple combination where $1-q$ of all nodes have $T=1$ and the
remaining nodes follow the exponential distribution
$\varrho(T)=\ee^{-(T-1)}$ for $T\in[1;\infty)$. By the same
approach as before, we obtain the equation for $\Omega^*$ in the
form $\ee^{1/\Omega^*}[1-q+q/(1-1/\Omega^*)]=2$ which yields
power-law exponents monotonically increasing from $2.44$ (for
$q=0$) to $4.18$ (for $q=1$). The reason for the exponent
decreasing as $q$ shrinks is that when $q$ is small, every node
with a potentially high exponentially-distributed $T$ value has
few able competitors during its life span and therefore it is
likely to acquire many links (more than for $q=1$). At the same
time, as $q$ decreases, the power-law tail contains smaller and
smaller fraction of all nodes and becomes less visible. This
example further demonstrates flexibility of the studied model
which is able to produce different kinds of behavior depending
on the input parameters.

It is easy to show that as long as $R_i(t)$ values decay faster
than $1/t$, the growth of $k_i(t)$ is sublinear and the
condensation phase observed in~\cite{BiaBar01b} is not
possible despite $T$ having an unlimited support. However, in
the system numerically studied in Fig.~\ref{fig:simulations},
deviations from the scale free distribution of node degree
appear when $\beta$ is small. This happens when the
characteristic lifetime of a node, $1/\beta$, is so long that
the decay cannot compensate for the unlimited support of
$\varrho(T)$. To get a qualitative estimate for the value of
$\beta$ when these deviations appear, we use the following
argument. If the final degree distribution is a power law with
exponent $\gamma$, we expect $\avg{k_{\mathrm{max}}}$ to grow as
$t^{1/(\gamma-1)}=\sqrt{t}$ (here we use that the number of
nodes equals $t$). When a node with a sufficiently high
relevance appears, the system can undergo a temporary
condensation phase where this node acquires a finite fraction of
links during its lifetime. To avoid a deviation from the power
law behavior, this lifetime must not be longer than
$\avg{k_{\mathrm{max}}}$, hence $\beta\lesssim1/\sqrt{t}$. As
$t$ goes to $\infty$, $\beta$ can be arbitrarily small and yet
no deviations appear. This confirms that in the thermodynamic
limit, the condensation phase does not realize in our model.

The key formula (\ref{k^F}) builds on the assumption that
fluctuations of $\Omega(t)$ are small enough, and the degree
distribution results hold if the effective lifetime of nodes
is long enough (a short-living node cannot acquire many links
regardless of its total relevance). These two assumptions are
in fact closely related: when the effective lifetime of nodes is
long, then at any time step there are many nodes competing for
the incoming link and the time fluctuations of $\Omega(t)$ are
hence small. To evaluate the effective life time of node $i$,
$\tau_i$, we use the participation number
\begin{equation}
\label{neff}
\tau_i:=\frac{\big(\sum_{t=1}^{\infty} R_i(t)\big)^2}
{\sum_{t=1}^{\infty} R_i(t)^2}\approx
T_i^2/\int_0^{\infty} R_i(t)^2\,\dd t.
\end{equation}
When $\tau_i\gg1$ for all nodes, $\Omega(t)$ fluctuates little.
Numerical simulations show that $\mathrm{var}(\Omega)$ is indeed
proportional to the effective life time for a wide range of
decay functions $R(t)$, confirming its relevance in the present
context. In conclusion, our analytical results are valid
when all the obtained conditions ($R_i(t)$ decreasing faster
than $1/t$, $\lvert\dot R_i(t)\rvert\ll R_i(t)$, and
$\tau_i\gg1$), are fulfilled.

To summarize, we studied a model of a growing network where
heterogeneous fitness (relevance) values and aging of nodes
(time decay) are combined. We showed that in contrast to models
where these two effects are considered in isolation, here we
obtain various realistic degree distributions for a wide range
of input parameters. We analyzed real citation data and showed
that they indeed support the hypothesis of coexisting node
heterogeneity and time decay. Even when our model is more
realistic than the preferential attachment alone, it neglects
several effects which might be of importance in various systems:
directed nature of the network, accelerating growth of the
network, gradual fragmentation of the network into related yet
independent fields, and others. Note that the very reason for
the exponential tail of the total fitness value $T$, though it
is crucial for the resulting degree distribution, is not
discussed here at all---yet we have empirical support for it in
our data. Also the case when the normalization $\Omega(t)$ in
Eq.~(\ref{basic}) does not have a stationary value (because
$\lim_{t\to\infty} R_i(t)>0$ or $\varrho(T)$ decays slower than
exponentially) is open. Finally, note that while we focused on
the degree distribution here, there are other network
characteristics---such as clustering coefficient and degree
correlations---that deserve further attention.

This work was supported by the EU FET-Open Grant 231200 (project
QLectives) and by the Swiss National Science Foundation Grant
200020-132253. We are grateful to the APS for providing us the
data set. We acknowledge helpful discussions with Yi-Cheng
Zhang, An Zeng, Juraj F\"oldes, Matou\v s Ringel and Yves Berset.


\begin{thebibliography}{99}
\bibitem{Yule25} G. U. Yule,
\emph{Phil. Trans. R. Soc. B} \textbf{213}, 21 (1925).

\bibitem{Price76} D. J. de S. Price,
\emph{J. of the Am. Soc. for Inf. Science} \textbf{27},
292 (1976).

\bibitem{Simon55} H. A. Simon,
\emph{Biometrika} \textbf{42}, 425 (1955).

\bibitem{BarAlb99} A. L. Barab\'asi, R. Albert,
\emph{Science} \textbf{286}, 509 (1999).

\bibitem{Newman03} M. E. J. Newman,
\emph{SIAM Review} \textbf{45}, 167 (2003).

\bibitem{Liu02} Z. Liu, Y.-C. Lai, N. Ye, P. Dasgupta,
\emph{Phys. Lett. A} \textbf{303}, 337 (2002).

\bibitem{KrReLe00} P. L. Krapivsky, S. Redner, F. Leyvraz,
Phys. Rev. Lett. 85, 4629 (2000).

\bibitem{BiaBar01} G. Bianconi, A.-L. Barab\'asi,
\emph{EPL} \textbf{54}, 436 (2001).

\bibitem{AlbBar02} R. Albert, A.-L. Barab\'asi,
\emph{Revs. of Mod. Phys.} \textbf{74}, 47 (2002).

\bibitem{Newman09} M. E. J. Newman,
\emph{EPL} \textbf{86}, 68001 (2009).

\bibitem{AdaHub00}  L. A. Adamic, B. A. Huberman,
\emph{Science} \textbf{287}, 2115 (2000).

\bibitem{Redner05} S. Redner,
\emph{Phys. Today} \textbf{58}, No. 6, 49 (2005).

\bibitem{AmScBaSt00} L. A. N. Amaral, A. Scala, M.
Barth\'el\'emy, H. E. Stanley,
\emph{Proc. Natl. Acad. Sci. U. S. A.} \textbf{97}, 11149 (2000).

\bibitem{LeJaLa05} S. Lehmann, A. D. Jackson, B. Lautrup,
\emph{EPL} \textbf{69}, 298 (2005).

\bibitem{DoMe00} S. N. Dorogovtsev, J. F. F. Mendes,
\emph{Phys. Rev. E} \textbf{62}, 1842 (2000).

\bibitem{BiaBar01b} G. Bianconi, A.-L. Barab\'asi,
\emph{Phys. Rev. Lett.} \textbf{86}, 5632 (2001).

\bibitem{JeNeBa03} H. Jeong, Z. N\'eda, A. L. Barab\'asi,
\emph{EPL} \textbf{61}, 567 (2003).

\bibitem{CSN09} A. Clauset, C. R. Shalizi, M. E. J. Newman,
\emph{SIAM Review} \textbf{51}, 661 (2009).

\bibitem{Zhu03} H. Zhu, X. Wang, J.-Y. Zhu,
\emph{Phys. Rev. E} \textbf{68}, 056121 (2003).

\bibitem{footnote1} Note that our definition of $X_j(t)$ is
consistent in the sense that when these values are used
in Eq.~(\ref{basic}), the expected degree increments
$C(t,\Delta t)P(i,t)$ equal the observed ones.

\bibitem{BoMaGo04} Katy B\"orner, J. T. Maru, R. L. Goldstone,
\emph{Proc. Natl. Acad. Sci. U. S. A.} \textbf{101}, 5266 (2004).

\bibitem{Mitz04} M. Mitzenmacher,
\emph{Internet Mathematics} \textbf{1}, 226 (2008).
\end{thebibliography}
\end{document}